\RequirePackage{lineno}
\documentclass[aps,twocolumn,showpacs,byrevtex,prl,reprint]{revtex4-1}

\usepackage{graphicx}
\usepackage{dcolumn}
\usepackage{bm}
\usepackage{rotating}
\usepackage{epstopdf}
\usepackage{color}
\usepackage{verbatim} 
\usepackage{multirow}
\usepackage[abs]{overpic}
\usepackage{amsmath}
\usepackage{amssymb}
\usepackage{subfigure}
\usepackage{xspace}

\usepackage[colorlinks,
            linkcolor=blue,
            anchorcolor=blue,
            citecolor=blue]{hyperref}

\newcommand{\PreserveBackslash}[1]{\let\temp=\\#1\let\\=\temp}
\newcolumntype{C}[1]{>{\PreserveBackslash\centering}p{#1}}
\newcolumntype{R}[1]{>{\PreserveBackslash\raggedleft}p{#1}}
\newcolumntype{L}[1]{>{\PreserveBackslash\raggedright}p{#1}}

\newcommand{\EE}{e^+e^-}

\newcommand{\GG}{\gamma\gamma}

\newcommand{\jpsi}{J/\psi}
\newcommand{\too}{\rightarrow}

\begin{document}
\graphicspath{{figure/}}
\DeclareGraphicsExtensions{.eps,.png,.ps}
\title{\boldmath Study of $\EE \too \pi^{0}X(3872)\gamma$ and search for $Z_c(4020)^{0}\too X(3872)\gamma$}
\author{
  \begin{small}
    \begin{center}
      M.~Ablikim$^{1}$, M.~N.~Achasov$^{10,c}$, P.~Adlarson$^{67}$, S. ~Ahmed$^{15}$, M.~Albrecht$^{4}$, R.~Aliberti$^{28}$, A.~Amoroso$^{66A,66C}$, M.~R.~An$^{32}$, Q.~An$^{63,49}$, X.~H.~Bai$^{57}$
% ...
\\
\vspace{0.2cm}
(BESIII Collaboration)\\
\vspace{0.2cm} {\it
$^{1}$ Institute of High Energy Physics, Beijing 100049, People's Republic of China\\
% ...
$^{4}$ Bochum Ruhr-University, D-44780 Bochum, Germany\\
% ...
$^{10}$ G.I. Budker Institute of Nuclear Physics SB RAS (BINP), Novosibirsk 630090, Russia\\
% ...
$^{15}$ Helmholtz Institute Mainz, Staudinger Weg 18, D-55099 Mainz, Germany\\
% ...
$^{28}$ Johannes Gutenberg University of Mainz, Johann-Joachim-Becher-Weg 45, D-55099 Mainz, Germany\\
% ...
$^{32}$ Liaoning Normal University, Dalian 116029, People's Republic of China\\
% ...
$^{49}$ State Key Laboratory of Particle Detection and Electronics, Beijing 100049, Hefei 230026, People's Republic of China\\
% ...
$^{57}$ University of Jinan, Jinan 250022, People's Republic of China\\
% ...
$^{63}$ University of Science and Technology of China, Hefei 230026, People's Republic of China\\
% ...
$^{66}$ University of Turin and INFN, (A)University of Turin, I-10125, Turin, Italy; (B)University of Eastern Piedmont, I-15121, Alessandria, Italy; (C)INFN, I-10125, Turin, Italy\\
$^{67}$ Uppsala University, Box 516, SE-75120 Uppsala, Sweden\\
% ...
$^{c}$ Also at the Novosibirsk State University, Novosibirsk, 630090, Russia\\
% ...
}\end{center}
    \vspace{0.4cm}
\end{small}
}
\affiliation{}


\begin{abstract}
Using data samples collected with the BESIII detector operating at the BEPCII storage ring at center-of-mass energies from 4.178 to 4.600 GeV, we study the process $\EE\rightarrow\pi^{0}X(3872)\gamma$ and search for $Z_c(4020)^{0}\too X(3872)\gamma$. We find no significant signal and set upper limits on $\sigma(\EE\rightarrow\pi^{0}X(3872)\gamma)\cdot\mathcal{B}(X(3872)\too\pi^{+}\pi^{-}J/\psi)$ and $\sigma(\EE\rightarrow\pi^{0}Z_c(4020)^{0})\cdot\mathcal{B}(Z_c(4020)^{0}\too
X(3872)\gamma)\cdot\mathcal{B}(X(3872)\too\pi^{+}\pi^{-}J/\psi)$ for each energy point at 90\% confidence level, which is of the order of several tenths pb.
\end{abstract}


\maketitle
\section{I. INTRODUCTION}

The recent discovery of several charmonium-like states has attracted great experimental and theoretical interests~\cite{pdg}. The charmonium-like states are also called $XYZ$ states where $X$ is the isospin-singlet state with $J^{PC}\neq1^{--}$, $Y$ is the isospin-singlet state with $J^{PC}=1^{--}$, and $Z$ is the isospin-triplet state~\cite{xyzstates}. The masses of these states are above the open-charm thresholds, and due to the unexpected resonance parameters and decay channels, these states can not be described by conventional quark models. Therefore, they are good candidates for exotic states, such as hybrids, tetraquarks, molecules, etc.~\cite{theory1, theory2, theory3}.

The first charmonium-like state $X(3872)$, which has been recently renamed as $\chi_{c1}(3872)$ by the Particle Data Group (PDG~\cite{pdg}), was observed by the Belle experiment in the process $B^{\pm}\too K^{\pm}X(3872)\too K^{\pm}\pi^{+}\pi^{-}J/\psi$~\cite{x38721}.
The $X(3872)$ is a rather narrow state with a mass that is consistent with $D^0\bar{D}^{*0}$ threshold. It decays through open-charm, radiative and isospin-violating pion emission decays, and is found to be an isospin singlet with $J^{PC} = 1^{++}$~\cite{pdg}.
Among these features, the extremely small mass difference between the $X(3872)$ and $D^{0}\bar{D}^{\ast0}$ threshold which we will denote as $\delta$, is of particular interest. Taking the values for the $D^0$, $D^{*0}$ and $X(3872)$ masses from the PDG~\cite{pdg}, $\delta$ is calculated to be $(-10\pm180)$ keV/$c^{2}$. Very recently, the LHCb reported a new measurement yielding $\delta =(70\pm120)$ keV/$c^{2}$~\cite{lhcb1, lhcb2}.  However the improved precision is still insufficient to tell whether the $X(3872)$ mass is above or below the $D^{0}\bar{D}^{\ast0}$ threshold.
Better knowledge of $\delta$ will be an important step towards a deeper understanding of the nature of the $X(3872)$~\cite{x3872-a, x3872-b}, and eventually of other related $XYZ$ states.
A completely new method to measure the $\delta$ value by measuring the $X(3872)\gamma$ line shape, which is sensitive to the $\delta$ value due to a triangle singularity, is proposed by Ref.~\cite{guo, ortega, guob}.
Here, the $X(3872)\gamma$ needs to be produced associated with another
positive $C$-parity neutral meson, e.g. $\EE\too\pi^0X(3872)\gamma$.
In principle, this method could be applied at the BESIII experiment,
based on the sizable data samples taken for $XYZ$ studies.
According to the theoretical prediction in Ref.~\cite{guob}, the cross section of the process $\EE\too\pi^0X(3872)\gamma$ is expected to be small. However, there could be other scenarios where the expected cross section is large.

Recently, the BESIII Collaboration reported an enhancement around 4.2 GeV for the $\EE\too\gamma X(3872)$ production cross sections~\cite{x3872-c}, which suggests a connection between $Y$ and $X$ states. BESIII also reported another connection, now between $Y$ and $Z$ states, with the observation of a $Y(4220)$ resonance in the process $\EE\too\pi^0 Z_c(3900)^0$~\cite{x3872-d}.
Those observations may indicate some common nature among the $XYZ$ states.
Therefore, it is important to search for possible connections between $Z$ and $X$ states. Establishing connections among $XYZ$ states may be a clue that can facilitate a better theoretical interpretation of these. One such connection~\cite{voloshin} could be a transition $Z_c(4020)^{0}\rightarrow X(3872)\gamma$ in the scenario where the $X(3872)$ is dominantly an $S$-wave $D^0\bar{D}^{*0}$ molecule and the $Z_c(4020)^{0}$ is an isotopic triplet of near-threshold $S$-wave $D^{*}\bar{D}^{*}$ resonances. Therefore, the search for the transition $Z_c(4020)^{0}\rightarrow X(3872)\gamma$ will help to quantitatively study the molecular picture of the $X(3872)$.
The $Z_c(4020)$ is observed in the $\EE\too\pi Z_c(4020)$ process, so the study of $\EE\too\pi^0X(3872)\gamma$ allows one to search for the transition $Z_c(4020)^{0}\rightarrow X(3872)\gamma$.

In this paper, we report the search for the reaction $\EE\rightarrow\pi^{0}X(3872)\gamma$ and  $Z_c(4020)^{0}\too X(3872)\gamma$ based on the data of twenty-three energy points recorded with the BESIII detector in the range of $4.178 \leq\sqrt{s}\leq4.600\,\rm{GeV}$. The $X(3872)$ state is reconstructed via $X(3872)\too\pi^{+}\pi^{-}J/\psi$, $J/\psi\too\ell^{+}\ell^{-}$ ($\ell=e$ or $\mu$).

\section{II. BESIII DETECTOR AND MONTE CARLO SIMULATION}
The BESIII detector is a magnetic spectrometer~\cite{besiii} located at the Beijing Electron
Positron Collider (BEPCII)~\cite{bepcii}. The
cylindrical core of the BESIII detector consists of a helium-based
multilayer drift chamber (MDC), a plastic scintillator time-of-flight
system (TOF), and a CsI(Tl) electromagnetic calorimeter (EMC),
which are all enclosed in a superconducting solenoidal magnet,
providing a 1.0~T magnetic field. The solenoid is supported by an
octagonal flux-return yoke with resistive plate chamber muon
identifier modules interleaved with steel. The acceptance of
charged particles and photons is 93\% over the $4\pi$ solid angle. The
charged-particle momentum resolution at $1~{\rm GeV}/c$ is
$0.5\%$, and the $\textrm{d}E/\textrm{d}x$ resolution is $6\%$ for the electrons
from Bhabha scattering. The EMC measures photon energies with a
resolution of $2.5\%$ ($5\%$) at $1$~GeV in the barrel (end cap)
region. The time resolution of the TOF barrel section is 68~ps, while that of the end cap section is 110~ps. The end cap TOF
system was upgraded in 2015 with multi-gap resistive plate chamber
technology, providing a time resolution of
60~ps~\cite{etof}. About 70\% of the data sample used here was taken after this upgrade.

Simulated data samples produced with the {\sc geant4}-based~\cite{geant4} Monte Carlo (MC) package, which
includes the geometric description of the BESIII detector and the
detector response, are used to determine the detection efficiency
and to estimate the background contributions.
The simulation includes the beam
energy spread and initial-state radiation (ISR) in the $e^+e^-$
annihilations modeled with the generator {\sc kkmc}~\cite{KKMC}.
The ISR production of vector charmonium(-like) states and the continuum processes are incorporated also in {\sc kkmc}~\cite{KKMC}. The known decay modes are modeled with {\sc evtgen}~\cite{ref:evtgen}, using branching fractions summarized and averaged by the
PDG~\cite{pdg}, and the remaining unknown decays
from the charmonium states are generated with {\sc lundcharm}~\cite{ref:lundcharm}. Final state radiation from charged final state particles is incorporated with the {\sc photos} package~\cite{photos}.

Signal MC samples for $\EE \too \pi^0X(3872)\gamma$ and $\EE \too \pi^0Z_c(4020)^0\too\pi^0X(3872)\gamma$ are generated according to phase space at each center-of-mass energy point, assuming that the cross section follows the function fit for the $\EE\too\pi^+\pi^-h_c$ line shape in Ref.~\cite{twobw}.
The event selection criteria and the detection efficiencies are determined and studied based on signal MC samples of $1\times10^{5}$ signal events generated for each value of $\sqrt{s}$.
Detection efficiencies are determined by the ratio of the reconstructed event yields (after the selection criteria) and the number of generated events.
Inclusive MC samples consisting of open charm production
processes are employed to investigate potential backgrounds.

\section{III. EVENT SELECTION}
For each charged track, the distance of closest approach to the interaction point (IP) is required to be within $10$ cm in the beam direction and within 1 cm in the plane perpendicular to the beam direction.
The polar angles ($\theta$) of the tracks with respect to the beam axis (ignoring the small crossing angle), must be within the fiducial volume of the MDC $(|\cos\theta|<0.93)$.
Photons are reconstructed from isolated showers in the EMC, which are at least $10^\circ$ away from the nearest charged track. The photon energy is required to be at least 25 MeV in the barrel region $(|\cos\theta|<0.80)$ or 50 MeV in the end cap region $(0.86<|\cos\theta|<0.92)$. To suppress electronic noise and energy depositions unrelated to the event, the EMC cluster timing from the reconstructed event start time is further required to satisfy $0\leq t \leq 700$ ns.

Since the reaction $\EE\too\pi^{0}X(3872)\gamma$ results in the final states $\gamma\gamma\gamma\pi^{+}\pi^{-}\ell^{+}\ell^{-}$, candidate events are required to have four tracks with zero net charge and at least three photons.
Tracks with momenta larger than 1.0~GeV/$c$ are assigned as leptons from the $J/\psi$ decay; otherwise, they are regarded as pions. Leptons from the $J/\psi$ decay with energy deposited in the EMC larger than 1.0~GeV are identified as electrons, while those with less than 0.4~GeV as muons. The $\pi^{0}$ candidates are reconstructed from photon pairs with invariant mass in the range $110<M_{\GG}<150\,{\rm MeV}/c^2$.

To reduce the background contributions and to improve the mass resolution,
a five-constraint (5C) kinematic fit is performed.  Four constraints come from the total initial four momentum of the colliding beams; the last one is from constraining the $M_{\GG}$ invariant mass to the nominal $\pi^0$ value~\cite{pdg}. If there is more than one combination in an event, the one with the smallest $\chi^{2}_{\text{5C}}$ is chosen.
Furthermore, the $\chi^{2}_{\text{5C}}$ is required to be less than 60. The $\jpsi$ is reconstructed via $\ell^{+}\ell^{-}$ decays, and the invariant mass of lepton pairs is required to be in the $\jpsi$ mass window $[3.080, 3.120]\,{\rm GeV}/c^{2}$.

\section{IV. BORN CROSS SECTION MEASUREMENT}

\subsection{IV.I $\EE\rightarrow\pi^{0}X(3872)\gamma$}
After applying the above requirements, the remaining background is mainly
coming from $e^{+}e^{-}\too\gamma_{ISR}\eta J/\psi$, $\eta\too\pi^{+}\pi^{-}\pi^{0}$
and $e^{+}e^{-}\too\gamma\omega J/\psi$, $\omega\too\pi^{+}\pi^{-}\pi^{0}$ events.
In order to veto these events, the $\pi^{+}\pi^{-}\pi^{0}$ invariant mass is required to be outside the $\eta$ and $\omega$ mass regions $[0.535, 0.560]\,{\rm GeV}/c^{2}$ and $[0.750, 0.810]\,{\rm GeV}/c^{2}$, respectively.
Besides the $\eta$ and $\omega$ backgrounds,
the $\pi^{+}\pi^{-}\psi(3686)$, $\psi(3686)\too\pi^{0}\pi^{0}\jpsi$ background is removed
by requiring the  $\pi^{+}\pi^{-}$ recoil mass to be outside
the $\psi(3686)$ mass region of $[3.670, 3.700]\,{\rm GeV}/c^2$.

\begin{figure*}[htbp]
\begin{center}
\begin{overpic}[width=0.9\textwidth]{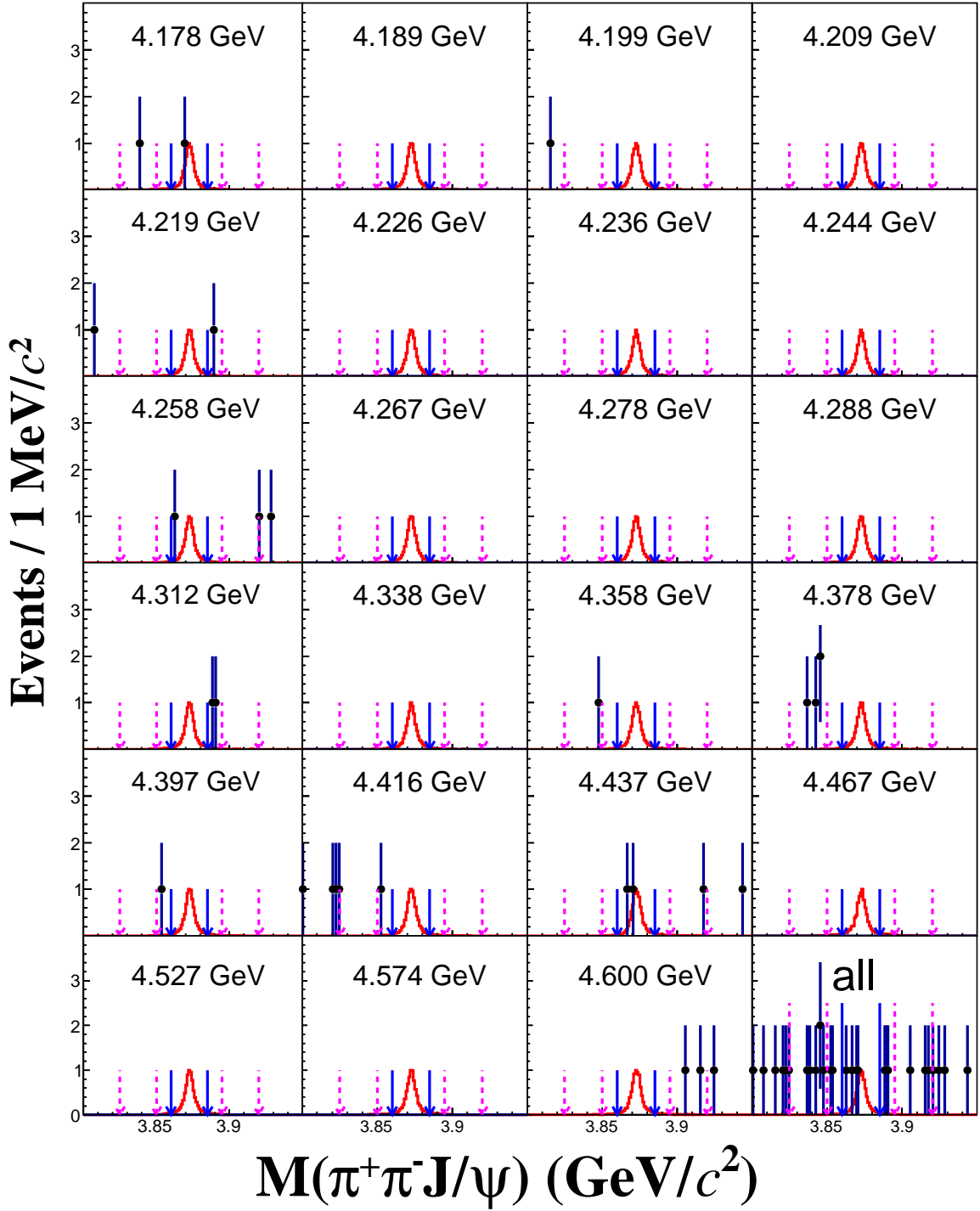}
\end{overpic}
\caption{The distribution of $M(\pi^{+}\pi^{-}J/\psi)$ for each energy point and the sum (all). Dots with error bars denote data, and the red histogram denotes the MC simulation of $\EE\rightarrow\pi^{0}X(3872)\gamma$. The blue solid lines mark the signal region of $X(3872)$, and the pink dashed lines mark the sideband regions of $X(3872)$.}
\label{fig:xx}
\end{center}
\end{figure*}

Figure~\ref{fig:xx} shows distributions of the $\pi^{+}\pi^{-}J/\psi$
invariant mass $M(\pi^{+}\pi^{-}J/\psi)$ for data and the MC samples
of $\EE\rightarrow\pi^{0}X(3872)\gamma$. The $X(3872)$ signal region
is taken as [3.860, 3.885] GeV/$c^{2}$, while the sideband regions
are set to be [3.825, 3.850] GeV/$c^{2}$ and [3.895, 3.920] GeV/$c^{2}$.
No significant $X(3872)$ signals are seen at any energies.
The signal yield is determined from the event yields in the $X(3872)$ signal and sideband regions.
The sideband yields are scaled by the ratio of the relevant mass-window widths
in order to predict the background expected in the signal region.
Upper limits on the number of signal events at the 90\% C.L. are calculated
by using a frequentist method~\cite{trolke1} with unbounded profile likelihood treatment of systematic uncertainties, which is implemented by the package {\sc trolke}~\cite{trolke2} in the {\sc root} framework~\cite{root}, where the signal and background obey Poisson statistics, and the efficiencies are Gaussian-distributed. The numerical results are summarized in Table~\ref{tab:prupperlimit}.

\begin{table*}[htbp]
  \centering
  \caption{The upper limits (calculated including the systematic uncertainties)
on $\sigma(\EE\rightarrow \pi^{0}X(3872)\gamma)\cdot\mathcal{B}(X(3872)\too\pi^{+}\pi^{-}J/\psi)$
and  $\sigma(\EE\rightarrow\pi^{0}Z_c(4020)^{0})\cdot\mathcal{B}(Z_c(4020)^{0}\too X(3872)\gamma)\cdot\mathcal{B}(X(3872)\too\pi^{+}\pi^{-}J/\psi)$
at the 90\% C.L. for each energy point, together with integrated luminosities $\mathcal{L}_{\rm int}$,
the number of events in signal region $N^\text{obs}$, the number of events in sideband region $N^\text{sb}$,
the number of signal events $N$ at the 90\% C.L., radiative correction factors 1+$\delta$(s), vacuum polarization
factors $\frac{1}{|1-\Pi|^{2}}$, and efficiencies without intermediate branching fractions $\epsilon$.
Here, $\sigma\cdot\mathcal{B}$ represents $\sigma(\EE\rightarrow \pi^{0}X(3872)\gamma)\cdot\mathcal{B}(X(3872)\too\pi^{+}\pi^{-}J/\psi)$ or $\sigma(\EE\rightarrow\pi^{0}Z_c(4020)^{0})\cdot\mathcal{B}(Z_c(4020)^{0}\too X(3872)\gamma)\cdot\mathcal{B}(X(3872)\too\pi^{+}\pi^{-}J/\psi)$.
The first values in brackets are for the process $\EE\rightarrow \pi^{0}X(3872)\gamma$,
and the second for the process $\EE\rightarrow\pi^{0}Z_c(4020)^{0}\too\pi^{0}X(3872)\gamma$.
The low efficiency at 4.467 GeV is caused by the cut on the $\pi^{+}\pi^{-}$ recoil mass.}
  \label{tab:prupperlimit}
  \begin{tabular}{c c c c c ccc c}
  \hline
  \hline
  \ \ \ $\sqrt{s}$ (GeV) \ \ \ & \ \ \ $\mathcal{L}_{\rm int}$(pb$^{-1}$) \ \ \ & \ \ \ $N^\text{obs}$ \ \ \ & \ \ \ $N^\text{sb}$ \ \ \ & \ \ \ $N$ \ \ \ & \ \ \ 1+$\delta$(s) \ \ \ & \ \ \ $\frac{1}{|1-\Pi|^{2}}$ \ \ \ & \ \ \ $\epsilon(\%)$ \ \ \ & \ \ \ $\sigma\cdot\mathcal{B}$(pb) \ \ \ \\
  \hline
  4.178 & 3195 & (1, 0) & (1, 1) & ($<3.27, <1.26$) & (0.70, 0.69) & 1.055 & (14.02, 13.97) & ($<0.08, <0.03$) \\
  4.189 & 527  & (0, 0) & (0, 2) & ($<2.00, <0.52$) & (0.70, 0.70) & 1.056 & (14.12, 14.02) & ($<0.31, <0.08$) \\
  4.199 & 526  & (0, 0) & (0, 0) & ($<2.00, <2.00$) & (0.70, 0.70) & 1.057 & (14.13, 14.24) & ($<0.31, <0.31$) \\
  4.209 & 517  & (0, 0) & (0, 0) & ($<2.00, <2.00$) & (0.71, 0.71) & 1.057 & (14.29, 13.75) & ($<0.31, <0.32$) \\
  4.219 & 515  & (0, 0) & (0, 0) & ($<2.00, <2.00$) & (0.72, 0.72) & 1.057 & (14.07, 13.74) & ($<0.31, <0.31$) \\
  4.226 & 1056 & (0, 0) & (0, 0) & ($<2.00, <2.00$) & (0.74, 0.74) & 1.057 & (14.51, 14.11) & ($<0.14, <0.15$) \\
  4.236 & 530  & (0, 0) & (0, 0) & ($<1.99, <2.00$) & (0.76, 0.76) & 1.056 & (14.50, 13.43) & ($<0.27, <0.30$) \\
  4.244 & 538  & (0, 0) & (0, 0) & ($<2.00, <2.00$) & (0.78, 0.78) & 1.056 & (14.03, 13.20) & ($<0.27, <0.29$) \\
  4.258 & 828  & (1, 0) & (0, 0) & ($<3.69, <2.01$) & (0.81, 0.81) & 1.054 & (14.00, 12.99) & ($<0.32, <0.19$) \\
  4.267 & 531  & (0, 0) & (0, 0) & ($<2.00, <2.01$) & (0.83, 0.83) & 1.053 & (13.78, 12.23) & ($<0.27, <0.30$) \\
  4.278 & 176  & (0, 0) & (0, 0) & ($<2.00, <2.02$) & (0.84, 0.84) & 1.053 & (13.44, 11.89) & ($<0.81, <0.93$) \\
  4.288 & 502  & (0, 0) & (0, 0) & ($<2.00, <2.01$) & (0.84, 0.84) & 1.053 & (13.29, 11.74) & ($<0.29, <0.33$) \\
  4.312 & 501  & (0, 0) & (0, 0) & ($<2.00, <2.02$) & (0.84, 0.84) & 1.052 & (13.35, 11.68) & ($<0.29, <0.33$) \\
  4.338 & 505  & (0, 0) & (0, 0) & ($<2.00, <2.02$) & (0.83, 0.83) & 1.051 & (13.76, 12.03) & ($<0.28, <0.32$) \\
  4.358 & 544  & (0, 0) & (1, 0) & ($<1.57, <2.02$) & (0.83, 0.83) & 1.051 & (14.11, 12.42) & ($<0.20, <0.29$) \\
  4.378 & 523  & (0, 0) & (4, 0) & ($<0.30, <2.02$) & (0.84, 0.84) & 1.052 & (14.06, 12.47) & ($<0.04, <0.30$) \\
  4.397 & 508  & (0, 0) & (0, 0) & ($<2.00, <2.02$) & (0.86, 0.86) & 1.052 & (13.60, 12.34) & ($<0.27, <0.30$) \\
  4.416 & 1044 & (0, 0) & (0, 0) & ($<2.00, <2.03$) & (0.90, 0.90) & 1.053 & (13.04, 12.10) & ($<0.13, <0.14$) \\
  4.437 & 570  & (2, 0) & (1, 1) & ($<4.93, <1.57$) & (0.97, 0.97) & 1.054 & (9.94, 11.47)  & ($<0.72, <0.20$) \\
  4.467 & 111  & (0, 0) & (0, 0) & ($<2.00, <2.03$) & (1.09, 1.09) & 1.055 & (5.25, 10.39)  & ($<2.53, <1.30$) \\
  4.527 & 112  & (0, 0) & (0, 0) & ($<2.00, <2.02$) & (1.38, 1.38) & 1.055 & (9.19, 8.56)   & ($<1.13, <1.23$) \\
  4.574 & 49   & (0, 0) & (0, 0) & ($<2.00, <2.03$) & (1.62, 1.62) & 1.055 & (8.11, 7.31)   & ($<2.50, <2.81$) \\
  4.600 & 587  & (0, 0) & (2, 0) & ($<1.15, <2.02$) & (1.76, 1.75) & 1.055 & (7.71, 7.06)   & ($<0.12, <0.22$) \\
  \hline
  \hline
  \end{tabular}
\end{table*}

\begin{figure*}[htbp]
\begin{center}
\begin{overpic}[width=0.43\textwidth]{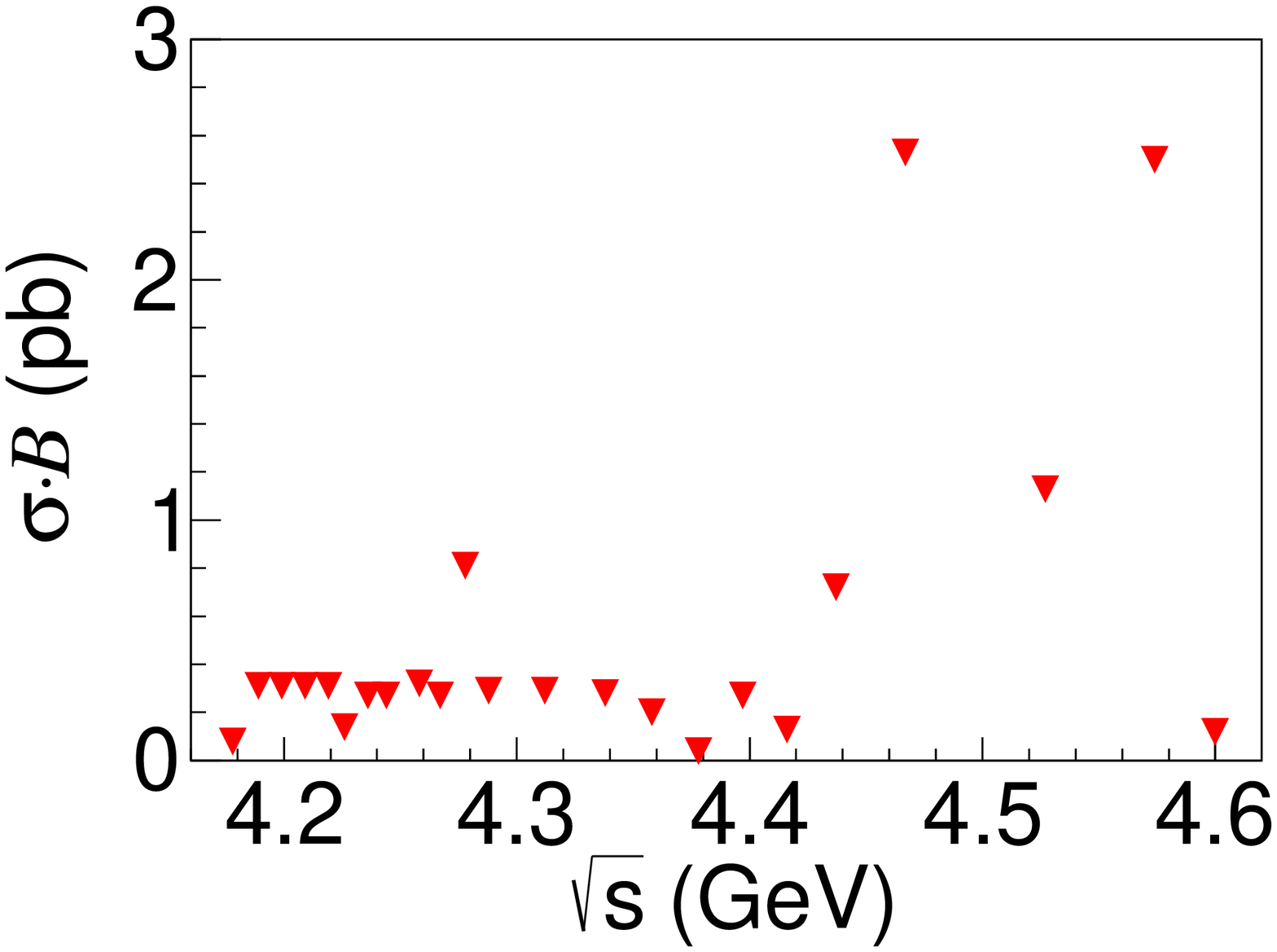}
\put(48,129){\LARGE{(a)}}
\end{overpic}
\begin{overpic}[width=0.43\textwidth]{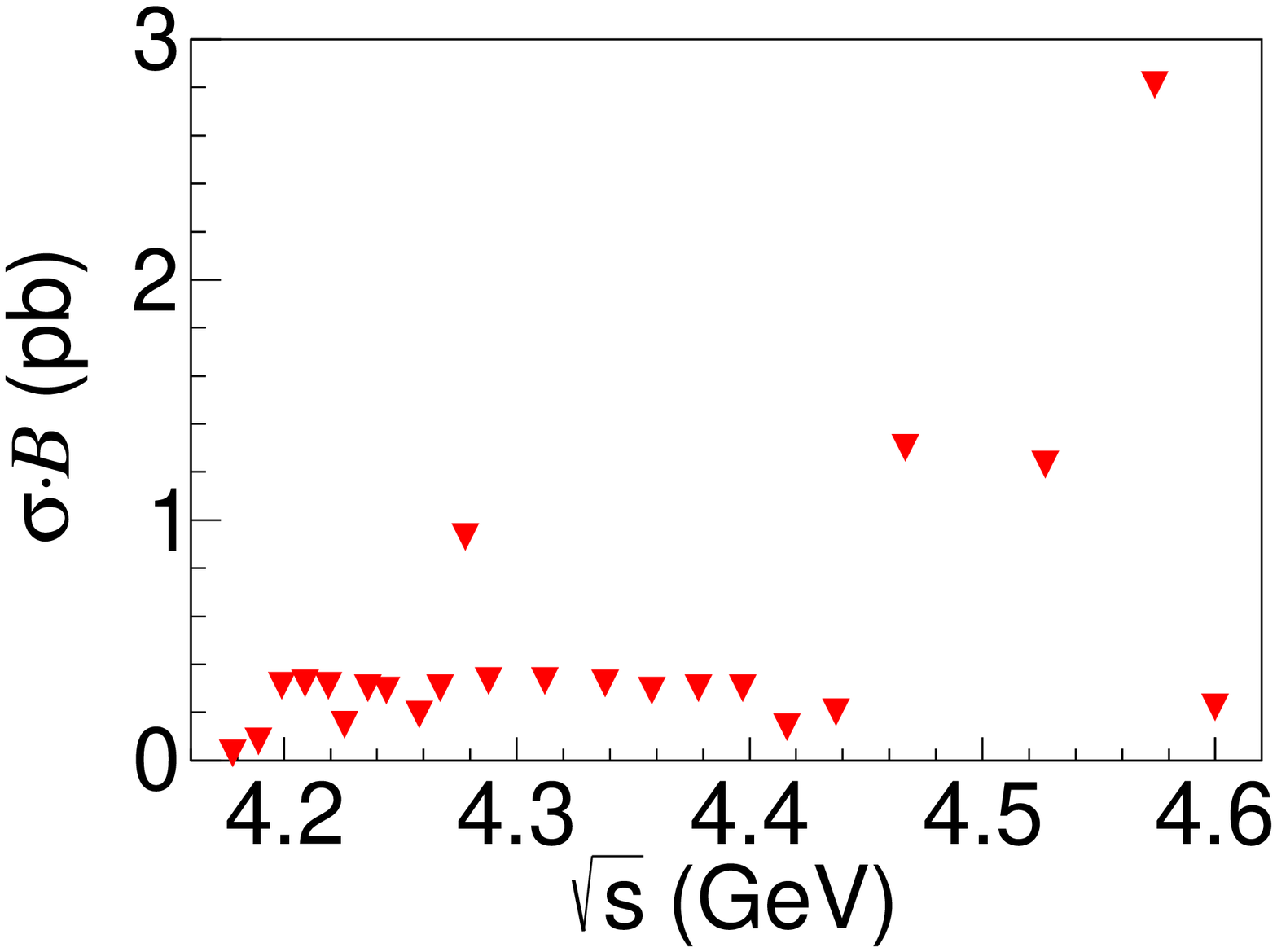}
\put(48,129){\LARGE{(b)}}
\end{overpic}
 \caption{The upper limits at the 90\% C.L. on  $\sigma(\EE\rightarrow\pi^{0}X(3872)\gamma)\cdot\mathcal{B}(X(3872)\too\pi^{+}\pi^{-}J/\psi)$ (a) and $\sigma(\EE\rightarrow\pi^{0}Z_c(4020)^{0})\cdot\mathcal{B}(Z_c(4020)^{0}\too X(3872)\gamma)\cdot\mathcal{B}(X(3872)\too\pi^{+}\pi^{-}J/\psi)$ (b) for each energy point. }
 \label{fig:upperx}
\end{center}
\end{figure*}

The Born cross section multiplied by the branching fraction
$\sigma(\EE\rightarrow \pi^{0}X(3872)\gamma)\cdot\mathcal{B}(X(3872)\too\pi^{+}\pi^{-}J/\psi)$
is calculated as:
\begin{equation}
\begin{aligned}
    &\sigma(\EE\rightarrow \pi^{0}X(3872)\gamma)\cdot\mathcal{B}(X(3872)\too\pi^{+}\pi^{-}J/\psi) =\\
    &\frac{N_{X(3872)}}{\epsilon \, \mathcal{L}_{\rm int} \, (1 + \delta (s)) \, \frac{1}{|1-\Pi|^{2}} \, {\mathcal{B}(J/\psi \too\ell^{+}\ell^{-})}\, {\mathcal{B}(\pi^{0} \too \gamma\gamma)}},
    \end{aligned}
\end{equation}
where
$N_{X(3872)}$ is the number of $X(3872)$ signal events,
$\epsilon$ is the detection efficiency (excluding intermediate branching fractions),
$\mathcal{L}_{\rm int}$ is the integrated luminosity~\cite{luminosity},
1+$\delta (s)$ is the ISR correction factor obtained from a quantum electrodynamics calculations~\cite{QED, KKMC},
$\frac{1}{|1-\Pi|^{2}}$ is vacuum polarization factor~\cite{vacuum}.
The corresponding upper limits for this cross section at the 90\% C.L.
for each energy point are listed in Table~\ref{tab:prupperlimit} and shown in Fig.~\ref{fig:upperx} (a).

Assuming the Born cross section of $\EE\rightarrow\pi^{0}X(3872)\gamma$ is constant at $4.178\leq\sqrt{s}\leq4.600$ GeV, the average Born cross section multiplied by the branching fraction
$\bar{\sigma}(\EE\rightarrow\pi^{0}X(3872)\gamma)\cdot\mathcal{B}(X(3872)\too\pi^{+}\pi^{-}J/\psi)$ for data is calculated as:
\begin{equation}
\begin{aligned}
    &\bar{\sigma}(\EE\rightarrow\pi^{0}X(3872)\gamma)\cdot\mathcal{B}(X(3872)\too\pi^{+}\pi^{-}J/\psi) =\\
    &\frac{N_{X(3872)}^{\text{total}}}{\sum\limits_i(\epsilon \, \mathcal{L}_{\rm int} \, (1 + \delta (s)) \, \frac{1}{|1-\Pi|^{2}})_i \, {\mathcal{B}(J/\psi \too\ell^{+}\ell^{-})}\, {\mathcal{B}(\pi^{0} \too \gamma\gamma)}},
    \end{aligned}
\end{equation}
where
$N_{X(3872)}^{\text{total}}$ is the total number of $X(3872)$ signal events,
and $i$ denotes each energy point.
The corresponding upper limit for the average Born cross section multiplied by the branching fraction
is determined to be 21.9 fb at the 90\% C.L.

\subsection{IV.II $Z_c(4020)^{0}\rightarrow X(3872)\gamma$}
\begin{figure*}[htbp]
\begin{center}
\begin{overpic}[width=0.9\textwidth]{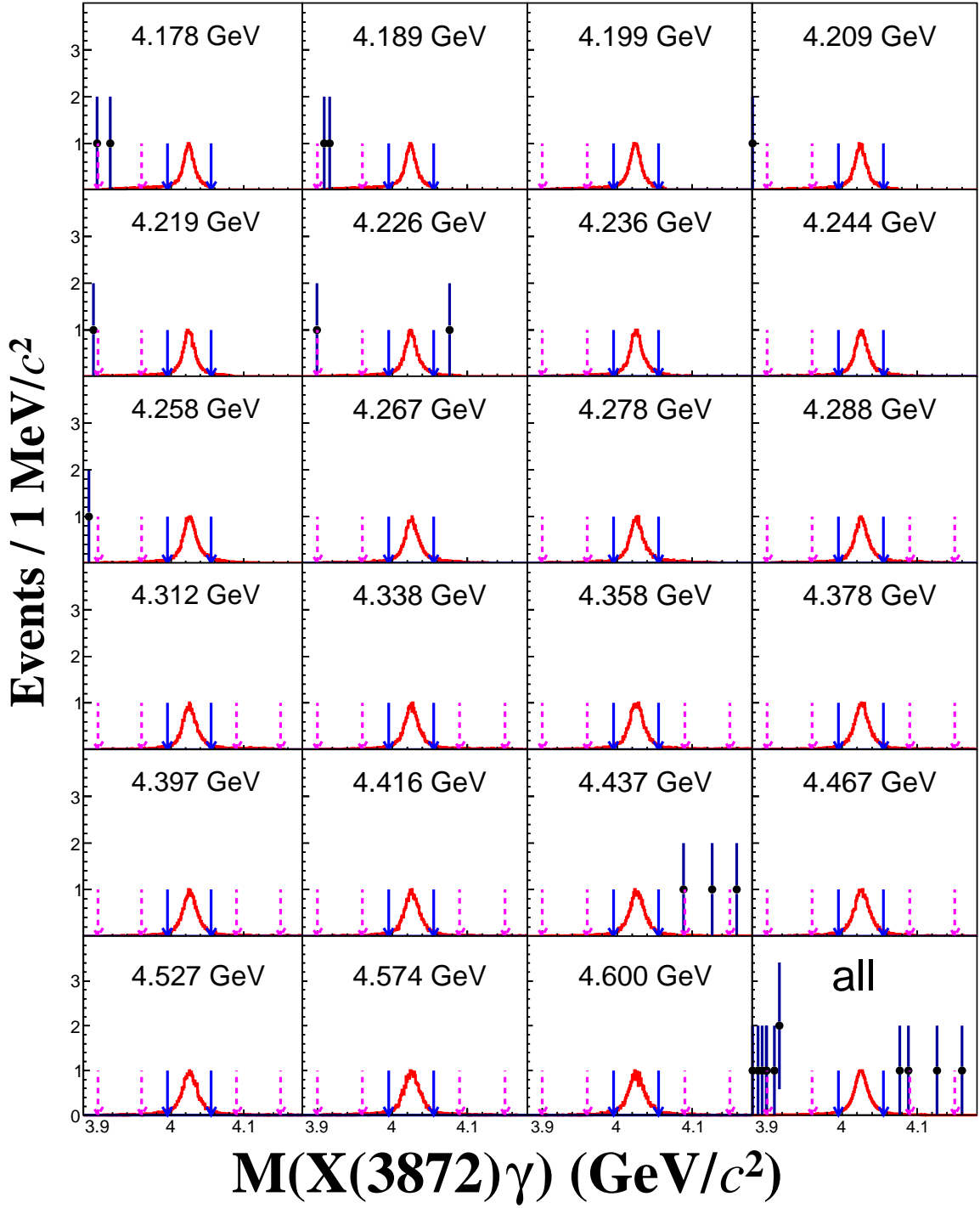}
\end{overpic}
\caption{The distribution of $M(X(3872)\gamma)$ for each energy point and the sum (all). Dots with error bars denote data, and the red histogram denotes the MC simulation of $\EE \too \pi^0Z_c(4020)^0\too\pi^0X(3872)\gamma$. The blue solid lines mark the signal region of $Z_c(4020)^{0}$, and the pink dashed lines mark the sideband regions of $Z_c(4020)^{0}$.}
\label{fig:xz}
\end{center}
\end{figure*}

The possible connection between $X$ and $Z$ charmonium-like states can be studied via the decay $Z_c(4020)^{0}\rightarrow X(3872)\gamma$.
In order to search for the process, the $X(3872)$ signal region is set to be [3.860, 3.885] GeV/$c^{2}$, which is the same as for the $\EE\rightarrow \pi^{0}X(3872)\gamma$ study.
As we do not observe any $X(3872)$ signal, there cannot be any $Z_c(4020)^{0}$ being produced in the given channel. Still, we provide corresponding upper limits, since a quantification might well be helpful in the understanding of the involved states.
After the requirement of the $X(3872)$ mass window
no significant $\eta$, $\omega$ and $\psi(3686)$ background remain.
Figure~\ref{fig:xz} shows the $X(3872)\gamma$ invariant mass distributions for data and MC samples of $\EE \too \pi^0Z_c(4020)^0\too\pi^0X(3872)\gamma$. No $Z_c(4020)^{0}$ candidates are found.
Therefore, the same method as before is employed to calculate the upper limits for this process.
For data samples taken above $\sqrt{s} = 4.280\,\rm{GeV}$,
the $Z_c(4020)^{0}$ signal region is set to be [3.995, 4.055] GeV/$c^{2}$,
and the sideband regions are set to be $[3.900, 3.960]\,{\rm GeV}/c^{2}$
and $[4.090, 4.150]\,{\rm GeV}/c^{2}$.
At lower energies, kinematics dictates that $Z_c(4020)^{0}$ candidates cannot have a mass
above $4.28-M_{\pi^0}=4.145\,{\rm GeV}/c^{2}$, where $M_{\pi^0}$ is $\pi^0$ nominal mass.
Accordingly, we use a single sideband region of [3.900, 3.960] GeV/$c^{2}$.

The Born cross section multiplied by branching fractions $\sigma(\EE\rightarrow\pi^{0}Z_c(4020)^{0})\cdot\mathcal{B}(Z_c(4020)^{0}\too X(3872)\gamma)\cdot\mathcal{B}(X(3872)\too\pi^{+}\pi^{-}J/\psi)$ is calculated with the following formula:
\begin{equation}
\begin{aligned}
    &\sigma(\EE\rightarrow\pi^{0}Z_c(4020)^{0})\cdot\mathcal{B}(Z_c(4020)^{0}\too X(3872)\gamma)\\
    &\cdot\mathcal{B}(X(3872)\too\pi^{+}\pi^{-}J/\psi) =\\
    &\frac{N_{Z_c(4020)^{0}}}{\epsilon \, \mathcal{L}_{\rm int} \, (1 + \delta (s)) \, \frac{1}{|1-\Pi|^{2}} \, {\mathcal{B}(J/\psi \too\ell^{+}\ell^{-})} \, {\mathcal{B}(\pi^{0} \too \gamma\gamma)}},
\end{aligned}
\end{equation}
where $N_{Z_c(4020)^{0}}$ is the number of $Z_c(4020)^{0}$ signal events.
The corresponding upper limits at the 90\% C.L.
for each energy are listed in Table~\ref{tab:prupperlimit} and shown in Fig.~\ref{fig:upperx} (b).

Assuming the Born cross section of $\EE\rightarrow\pi^{0}Z_c(4020)^{0}$ is constant at $4.178\leq\sqrt{s}\leq4.600$ GeV, the average Born cross section multiplied by branching fractions $\bar{\sigma}(\EE\rightarrow\pi^{0}Z_c(4020)^{0})\cdot\mathcal{B}(Z_c(4020)^{0}\too X(3872)\gamma)\cdot\mathcal{B}(X(3872)\too\pi^{+}\pi^{-}J/\psi)$ for data is calculated with the following formula:
\begin{equation}
\begin{aligned}
    &\bar{\sigma}(\EE\rightarrow\pi^{0}Z_c(4020)^{0})\cdot\mathcal{B}(Z_c(4020)^{0}\too X(3872)\gamma)\\
    &\cdot\mathcal{B}(X(3872)\too\pi^{+}\pi^{-}J/\psi) =\\
    &\frac{N_{Z_c(4020)^{0}}^{\text{total}}}{\sum\limits_i(\epsilon \, \mathcal{L}_{\rm int} \, (1 + \delta (s)) \, \frac{1}{|1-\Pi|^{2}})_i \, {\mathcal{B}(J/\psi \too\ell^{+}\ell^{-})} \, {\mathcal{B}(\pi^{0} \too \gamma\gamma)}},
\end{aligned}
\end{equation}
where $N_{Z_c(4020)^{0}}^{\text{total}}$ is the total number of $Z_c(4020)^{0}$ signal events.
The corresponding upper limit for the average Born cross section multiplied by the branching fraction
is determined to be 1.6 fb at the 90\% C.L.

\section{V. SYSTEMATIC UNCERTAINTY ESTIMATION}
The systematic uncertainties of $\sigma(\EE\rightarrow\pi^{0}X(3872)\gamma)\cdot\mathcal{B}(X(3872)\too\pi^{+}\pi^{-}J/\psi)$ and $\sigma(\EE\rightarrow\pi^{0}Z_c(4020)^{0})\cdot\mathcal{B}(Z_c(4020)^{0}\too X(3872)\gamma)\cdot\mathcal{B}(X(3872)\too\pi^{+}\pi^{-}J/\psi)$ originate from the luminosity measurement, the tracking efficiency, the photon detection efficiency, the kinematic fit, the $J/\psi$ mass window, the $X(3872)$ mass window, the $Z_c(4020)^{0}$ parameters, the line shape, the generator model, the ISR correction, and the input branching fractions.

The integrated luminosity at each point has been measured with a precision of $1.0\%$ using the Bhabha process~\cite{luminosity}.

The uncertainty from the tracking efficiency is $1.0\%$ per track~\cite{omegachic0} and the uncertainty in photon detection efficiency is $1.0\%$ per photon~\cite{photon}.

The uncertainty due to the kinematic fit requirements is estimated by correcting the helix parameters of charged tracks according to the method described in Ref.~\cite{helix}. The difference between detection efficiencies obtained from MC samples with and without this correction is taken as the uncertainty.

The uncertainty for the $J/\psi$ mass window is estimated using the control sample of $\EE \too \gamma_{\rm ISR}\psi(3686), \psi(3686) \too \pi^{+}\pi^{-}J/\psi$. The difference of the efficiency between data and MC simulation is found to be 1.6$\%$~\cite{jpsimasswindow}, which is taken as the uncertainty.

The uncertainty from the $X(3872)$ mass window is estimated by changing the window range by $\pm$10\%, and the largest efficiency change is taken as the uncertainty.

The uncertainties arising from the $Z_c(4020)^{0}$ mass and width are estimated by changing them by one standard deviation values~\cite{pdg} while generating the signal MC. The largest efficiency difference relative to the nominal one is taken as the uncertainty.

The line shape affects the ISR correction factor and the efficiency.
No obvious signal was found for our $Z_c(4020)^{0}$ search, so we use the line shape from
$\EE\too\pi^+\pi^-h_c$ in Ref.~\cite{twobw} as the input line shape to get the nominal results. To get the uncertainty introduced by the line shape, we change it to a Breit-Wigner function describing the $\psi(4230)$ or $\psi(4415)$, with the masses and widths fixed to the values from PDG~\cite{pdg}. The largest difference of the final result is taken as a systematic uncertainty.

For the systematic uncertainty from the MC simulation describing the process $\EE\too\pi^{0}X(3872)\gamma$, we use the three-body phase space MC simulation to get the nominal efficiency, then change to the $\EE\too\pi^{0}Z_c(4020)^{0}\too\pi^0 X(3872)\gamma$.
The difference on the detection efficiency with and without the intermediate resonant state is taken as the uncertainty due to the MC generator model.

The systematic uncertainty from the MC simulation describing
the process $\EE\too\pi^{0}Z_c(4020)^{0}\too\pi^0 X(3872)\gamma$
is estimated by varying the distribution of the $\pi^0$ polar angle $\theta$.
The nominal efficiency is determined assuming a flat distribution in $\cos{\theta}$.
A conservative estimate of the systematic uncertainty is obtained using
alternative MC samples with angular distributions of $1\pm\cos^{2}{\theta}$.
The largest change of efficiency is taken as the uncertainty due to the MC generator model.

The ISR correction factor is obtained from quantum electrodynamics calculations~\cite{QED, KKMC}. We also analyze MC samples with and without ISR effects considered to get the ISR correction factor, the difference of the two results is taken as the systematic uncertainty on the ISR correction factor.

As uncertainties introduced by the branching fractions of $J/\psi \too\ell^{+}\ell^{-} $ and $\pi^{0} \too \gamma\gamma$ we use those quoted by the PDG~\cite{pdg}.

Table~\ref{tab:summererror} summarizes all the systematic uncertainties related to $\sigma(\EE\rightarrow\pi^{0}X(3872)\gamma)\cdot\mathcal{B}(X(3872)\too\pi^{+}\pi^{-}J/\psi)$ and $\sigma(\EE\rightarrow\pi^{0}Z_c(4020)^{0})\cdot\mathcal{B}(Z_c(4020)^{0}\too X(3872)\gamma)\cdot\mathcal{B}(X(3872)\too\pi^{+}\pi^{-}J/\psi)$ for each center-of-mass energy. The total systematic uncertainty for each energy point
is calculated as the quadratic sum of the individual uncertainties, assuming them to be uncorrelated.

\begin{table*}[htbp]
  \centering
  \caption{Summary of relative systematic uncertainties (\%) associated with luminosity($\mathcal{L}_{\rm int}$), tracking efficiency (Tracks), photon detection efficiency (Photons), kinematic fitting ($\chi^{2}_{5C}$), $J/\psi$ mass window ($J/\psi$), $X(3872)$ mass window ($X(3872)$), $Z_c(4020)^{0}$ parameters ($Z_c(4020)^{0}$), line shape (Line shape), generator model (Generator), ISR correction factor (ISR) and branching fraction ($\mathcal{B}$). The first values in brackets are for the process $\EE\rightarrow \pi^{0}X(3872)\gamma$, and the second for the process $\EE\rightarrow\pi^{0}Z_c(4020)^{0}\too\pi^{0}X(3872)\gamma$. A dash indicates that a systematic effect is not applicable.}
  \label{tab:summererror}
  \begin{tabular}{ccccccccccccc}
  \hline
  \hline
  $\sqrt{s}$ (GeV) & $\mathcal{L}_{\rm int}$ & Tracks & Photons & $\chi^{2}_{5C}$ & $J/\psi$ & $X(3872)$ & $Z_c(4020)^{0}$ & Line shape & Generator & ISR & $\mathcal{B}$ & Sum \\
  \hline
  4.178 & 1.0 & 4.0 & 3.0 & (2.6, 2.0) & 1.6 &  1.3 & (-, 4.2) & (5.6, 6.0) & (7.4, 2.4)  & (0.7, 0.7) & 0.4 & (11.1, 9.7) \\
  4.189 & 1.0 & 4.0 & 3.0 & (2.8, 2.1) & 1.6 &  1.3 & (-, 3.5) & (6.4, 6.8) & (5.7, 3.9)  & (0.7, 0.6) & 0.4 & (10.6, 10.4) \\
  4.199 & 1.0 & 4.0 & 3.0 & (2.1, 2.2) & 1.6 &  1.3 & (-, 4.3) & (6.7, 4.7) & (7.5, 3.9)  & (0.5, 0.5) & 0.4 & (11.7, 9.6) \\
  4.209 & 1.0 & 4.0 & 3.0 & (2.0, 2.1) & 1.6 &  1.4 & (-, 4.1) & (4.9, 6.3) & (3.5, 6.6)  & (0.2, 0.3) & 0.4 & (8.4, 11.6) \\
  4.219 & 1.0 & 4.0 & 3.0 & (2.4, 2.6) & 1.6 &  1.5 & (-, 5.7) & (4.3, 3.5) & (5.2, 7.7)  & (0.1, 0.1) & 0.4 & (9.1, 11.9) \\
  4.226 & 1.0 & 4.0 & 3.0 & (2.3, 2.1) & 1.6 &  1.5 & (-, 5.6) & (1.5, 2.3) & (5.1, 7.1)  & (0.1, 0.1) & 0.4 & (8.0, 11.1) \\
  4.236 & 1.0 & 4.0 & 3.0 & (2.3, 2.1) & 1.6 &  1.5 & (-, 6.3) & (2.1, 2.2) & (1.3, 9.2)  & (0.1, 0.1) & 0.4 & (6.5, 12.8) \\
  4.244 & 1.0 & 4.0 & 3.0 & (2.1, 2.3) & 1.6 &  1.3 & (-, 4.6) & (4.4, 1.3) & (3.3, 9.7)  & (0.2, 0.1) & 0.4 & (8.1, 12.4) \\
  4.258 & 1.0 & 4.0 & 3.0 & (2.3, 2.6) & 1.6 &  1.6 & (-, 5.8) & (6.4, 4.4) & (3.3, 9.5)  & (0.2, 0.3) & 0.4 & (9.4, 13.5) \\
  4.267 & 1.0 & 4.0 & 3.0 & (2.0, 2.0) & 1.6 &  1.3 & (-, 5.9) & (5.7, 6.8) & (0.2, 12.9) & (0.2, 0.1) & 0.4 & (8.2, 16.8) \\
  4.278 & 1.0 & 4.0 & 3.0 & (2.4, 2.1) & 1.6 &  1.5 & (-, 5.6) & (6.9, 7.2) & (1.0, 13.6) & (0.2, 0.1) & 0.4 & (9.2, 17.4) \\
  4.288 & 1.0 & 4.0 & 3.0 & (2.1, 2.2) & 1.6 &  1.4 & (-, 5.9) & (7.9, 6.0) & (1.3, 13.5) & (0.2, 0.1) & 0.4 & (10.0, 17.0) \\
  4.312 & 1.0 & 4.0 & 3.0 & (2.8, 2.1) & 1.6 &  1.3 & (-, 6.6) & (5.8, 5.9) & (2.2, 15.4) & (0.1, 0.2) & 0.4 & (8.8, 18.7) \\
  4.338 & 1.0 & 4.0 & 3.0 & (2.1, 2.2) & 1.6 &  1.5 & (-, 5.4) & (7.0, 5.6) & (2.9, 16.0) & (0.1, 0.1) & 0.4 & (9.6, 18.8) \\
  4.358 & 1.0 & 4.0 & 3.0 & (2.0, 1.8) & 1.6 &  1.2 & (-, 5.6) & (7.6, 6.1) & (2.8, 15.5) & (0.1, 0.1) & 0.4 & (10.0, 18.5) \\
  4.378 & 1.0 & 4.0 & 3.0 & (2.1, 1.8) & 1.6 &  1.5 & (-, 6.0) & (7.0, 3.9) & (3.5, 14.9) & (0.1, 0.1) & 0.4 & (9.8, 17.5) \\
  4.397 & 1.0 & 4.0 & 3.0 & (2.1, 1.8) & 1.6 &  1.3 & (-, 7.4) & (5.3, 5.5) & (5.3, 16.4) & (0.1, 0.1) & 0.4 & (9.5, 19.7) \\
  4.416 & 1.0 & 4.0 & 3.0 & (1.8, 1.7) & 1.6 &  1.3 & (-, 7.1) & (4.4, 5.5) & (4.3, 19.0) & (0.1, 0.1) & 0.4 & (8.5, 21.8) \\
  4.437 & 1.0 & 4.0 & 3.0 & (2.1, 2.0) & 1.6 &  1.4 & (-, 7.1) & (3.1, 1.2) & (5.9, 17.4) & (0.1, 0.2) & 0.4 & (8.9, 19.7) \\
  4.467 & 1.0 & 4.0 & 3.0 & (2.4, 2.2) & 1.6 &  1.5 & (-, 7.3) & (3.7, 5.5) & (9.8, 18.2) & (0.1, 0.1) & 0.4 & (12.1, 21.2) \\
  4.527 & 1.0 & 4.0 & 3.0 & (2.6, 1.5) & 1.6 &  1.5 & (-, 7.4) & (5.8, 2.1) & (5.5, 17.8) & (0.2, 0.3) & 0.4 & (10.1, 20.2) \\
  4.575 & 1.0 & 4.0 & 3.0 & (2.3, 1.5) & 1.6 &  1.4 & (-, 7.2) & (4.1, 3.0) & (7.4, 20.1) & (0.5, 0.5) & 0.4 & (10.4, 22.3) \\
  4.600 & 1.0 & 4.0 & 3.0 & (2.3, 1.9) & 1.6 &  1.4 & (-, 6.6) & (0.4, 1.1) & (8.1, 16.2) & (0.5, 0.6) & 0.4 & (10.1, 18.5) \\
  \hline
  \hline
  \end{tabular}\\
\end{table*}

\section{VI. SUMMARY}
Using data samples collected at the center-of-mass energies between 4.178 and 4.600 GeV, the processes $\EE\rightarrow\pi^{0}X(3872)\gamma$  and $Z_c(4020)^{0}\rightarrow X(3872)\gamma$ are investigated.
In neither of the two processes are significant signals observed.
Upper limits at the 90\% C.L. on the cross sections multiplied by the branching fractions, $\sigma(\EE\rightarrow\pi^{0}X(3872)\gamma)\cdot\mathcal{B}(X(3872)\too\pi^{+}\pi^{-}J/\psi)$ and $\sigma(\EE\rightarrow\pi^{0}Z_c(4020)^{0})\cdot\mathcal{B}(Z_c(4020)^{0}\too
X(3872)\gamma)\cdot\mathcal{B}(X(3872)\too\pi^{+}\pi^{-}J/\psi)$, are reported for each energy point. The average cross sections multiplied by branching fractions are also determined. The measured results of the process $\EE\rightarrow\pi^{0}X(3872)\gamma$
are not in conflict with the theoretical expectation of about 0.1 fb~\cite{guob}.
A three orders of magnitude increase in statistics is needed to test these models.

Using the experimental results on the $\sigma(\EE\rightarrow\pi^{0}Z_c(4020)^{0})\cdot\mathcal{B}(Z_c(4020)^{0}\too
(D^{\ast}\bar{D}^{\ast})^{0})$ at $\sqrt{s}=4.226$ and $4.258$ GeV~\cite{zc40202}, the ratio $\frac{\mathcal{B}(Z_c(4020)^{0}\too X(3872)\gamma)\cdot\mathcal{B}(X(3872)\too\pi^{+}\pi^{-}J/\psi)}{\mathcal{B}(Z_c(4020)^{0}\too(D^{\ast}\bar{D}^{\ast})^{0})}$ is determined to be less than 0.15\% at the 90\% C.L.
The ratio does not contradict the prediction reported in Ref.~\cite{voloshin} based on the molecular picture.
Since no significant $\EE\rightarrow\pi^{0}X(3872)\gamma$ signals are observed, we cannot study the lineshape as proposed in Ref.~\cite{guo, ortega}; this may be achieved at future super tau-charm facilities~\cite{supertaocharm1, supertaocharm2}.

\section{ACKNOWLEDGMENTS}

The BESIII collaboration thanks the staff of BEPCII and the IHEP computing center for their strong support. This work is supported in part by National Key Research and Development Program of China under Contracts Nos. 2020YFA0406300, 2020YFA0406400; National Natural Science Foundation of China (NSFC) under Contracts Nos. 11905179, 11625523, 11635010, 11735014, 11822506, 11835012, 11935015, 11935016, 11935018, 11961141012; the Chinese Academy of Sciences (CAS) Large-Scale Scientific Facility Program; Joint Large-Scale Scientific Facility Funds of the NSFC and CAS under Contracts Nos. U1732263, U1832207; CAS Key Research Program of Frontier Sciences under Contracts Nos. QYZDJ-SSW-SLH003, QYZDJ-SSW-SLH040; 100 Talents Program of CAS; INPAC and Shanghai Key Laboratory for Particle Physics and Cosmology; ERC under Contract No. 758462; European Union Horizon 2020 research and innovation programme under Contract No. Marie Sklodowska-Curie grant agreement No 894790; Nanhu Scholars Program for Young Scholars of Xinyang Normal University; German Research Foundation DFG under Contracts Nos. 443159800, Collaborative Research Center CRC 1044, FOR 2359, FOR 2359, GRK 214; Istituto Nazionale di Fisica Nucleare, Italy; Ministry of Development of Turkey under Contract No. DPT2006K-120470; National Science and Technology fund; Olle Engkvist Foundation under Contract No. 200-0605; STFC (United Kingdom); The Knut and Alice Wallenberg Foundation (Sweden) under Contract No. 2016.0157; The Royal Society, UK under Contracts Nos. DH140054, DH160214; The Swedish Research Council; U. S. Department of Energy under Contracts Nos. DE-FG02-05ER41374, DE-SC-0012069.

\end{document}